\documentclass[12pt,preprint]{aastex}
\usepackage{graphicx} 
\usepackage{color} 

\shortauthors{Ferraro et al.}

\def\ltsima{$\; \buildrel < \over \sim \;$}
\def\gtsima{$\; \buildrel > \over \sim \;$}
\def\lsim{\lower.5ex\hbox{\ltsima}}
\def\gsim{\lower.5ex\hbox{\gtsima}}

\begin{document}

\title{The Hubble Space Telescope UV Legacy Survey of Galactic
  Globular Clusters - XV. The dynamical clock: reading cluster
  dynamical evolution from the segregation level of blue straggler
  stars}

\author{F.R. Ferraro$^{1,2}$, B. Lanzoni$^{1,2}$, S. Raso$^{1,2}$,
  D. Nardiello$^3$, E. Dalessandro$^2$, E. Vesperini$^4$,
  G. Piotto$^3$, C. Pallanca$^{1,2}$, G. Beccari$^5$, A. Bellini$^6$,
  M. Libralato$^6$, J. Anderson$^6$, A. Aparicio$^{7,8}$,
  L.R. Bedin$^{9}$, S. Cassisi$^{10}$, A.P. Milone$^3$,
  S. Ortolani$^3$, A. Renzini$^9$, M. Salaris$^{11}$, R.P. van der
  Marel$^{6,12}$ }

\affil{
\textsuperscript{1} Dept. of Physics and Astronomy, University of Bologna, Via Gobetti 93/2, Bologna, Italy\\
\textsuperscript{2} INAF Osservatorio di Astrofisica e Scienza dello Spazio di Bologna, Via Gobetti 93/3, Bologna, Italy\\
\textsuperscript{3} Dept. of Physics and Astronomy Galileo Galilei, University of Padova, Vicolo dell'Osservatorio 3, I-35122 Padova, Italy\\
\textsuperscript{4} Dept. of Astronomy, Indiana University, Bloomington, IN, 47401, USA\\
\textsuperscript{5} European Southern Observatory, Karl-Schwarzschild-Strasse 2, 85748  Garching bei M\"unchen, Germany\\
\textsuperscript{6} Space Telescope Science Institute, 3700 San Martin Drive, Baltimore, MD 21218, USA.\\
\textsuperscript{7} Instituto de Astrofisica de Canarias, E-38200 La Laguna, Tenerife, Canary Islands, Spain\\
\textsuperscript{8} Department of Astrophysics, University of La Laguna, E-38200 La Laguna, Tenerife, Canary Islands, Spain\\
\textsuperscript{9} INAF-Osservatorio Astronomico di Padova, Vicolo dell'Osservatorio 5, I-35122 Padova, Italy\\ 
\textsuperscript{10} Osservatorio Astronomico di Teramo, Via Mentore Maggini s.n.c., I-64100 Teramo, Italy\\
\textsuperscript{11} Astrophysics Research Institute, Liverpool John Moores University, Liverpool Science Park, IC2 Building,
146 Brownlow Hill, Liverpool L3 5RF, UK\\ 
\textsuperscript{12} Center for Astrophysical Sciences, Department of Physics \& Astronomy,
Johns Hopkins University, Baltimore, MD 21218, USA
}

\date{13 April 2018}

\begin{abstract}
The parameter $A^+$, defined as the area enclosed between the
cumulative radial distribution of blue straggler stars (BSSs) and that
of a reference population, is a powerful indicator of the level of BSS
central segregation.  As part of the Hubble Space Telescope UV Legacy
Survey of Galactic globular clusters (GCs), here we present the BSS
population and the determination of $A^+$ in 27 GCs observed out to
about one half-mass radius.  In combination with 21 additional
clusters discussed in a previous paper this provides us with a global
sample of 48 systems (corresponding to $\sim 32\%$ of the Milky Way GC
population), for which we find a strong correlation between $A^+$ and
the ratio of cluster age to the current central relaxation time. 
  Tight relations have been found also with the core radius and the
  central luminosity density, which are expected to change with the
  long-term cluster dynamical evolution. An interesting relation is
emerging between $A^+$ and the ratio of the BSS velocity dispersion
relative to that of main sequence turn-off stars, which measures the
degree of energy equipartition experienced by BSSs in the cluster.
These results provide further confirmation that BSSs are invaluable
probes of GC internal dynamics and $A^+$ is a powerful dynamical
clock.
\end{abstract}


\section{Introduction}
\label{intro}
Blue straggler stars (BSSs) appear as a sparse group of stars lying
along an extrapolation of the main sequence (MS), at brighter
magnitudes and bluer colors than the MS turnoff (MS-TO), in the
color-magnitude diagram (CMD) of old stellar populations (e.g.,
\citealp{sandage53, ferraro+92, ferraro+93, ferraro+97, ferraro+04,
  ferraro+06a, lanzoni+07a, lanzoni+07b, leigh07, moretti08,
  dalessandro+08, beccari+11, simunovic+16}).  Their location in the
CMD and a growing set of additional observational evidence
\citep{shara97, gilliland98, ferraro+06b, fiorentino+14} suggest that
they are more massive than any cluster star with active thermonuclear
burnings.  In the absence of any recent star-formation event, the
origin of BSSs must be sought by exploring mechanisms that are able to
increase the initial mass of single stars.  Two main BSS formation
channels have been identified so far: (1) mass-transfer in binary
systems, possibly up to the complete coalescence of the two companions
(\citealt{mccrea1964}), and (2) stellar mergers resulting from direct
collisions (\citealt{hills+1976}). The relative importance of each
process likely depends on the physical properties of the parent
cluster \citep[e.g.,][]{davies04, sollima08, chen09, knigge09,
  ferraro+09, leigh13, chatterjee13_bss, sills13}, and there is no way
at present to properly quantify this.  Moreover, a few indications
suggest that both channels can be active with comparable efficiency
within the same cluster \citep{ferraro+09, dalessandro+13_n362,
  simunovic+14, xin15}.  Irrespective of their formation mechanism,
BSSs are a population of heavy objects ($\sim 1$-$1.6 M_\odot$)
orbiting in a sea of lighter stars (the average stellar mass in an old
GC is $\langle m\rangle\sim 0.3 M_\odot$) with a density distribution
that varies by several orders of magnitude between the center and the
periphery of the host cluster (e.g., \citealp{mclvdm05, miocchi+13,
  baumgardt17}).  For this reason BSSs can be used as powerful {\it
  gravitational probe particles} to investigate key dynamical
processes (such as mass segregation) characterizing the dynamical
evolution of star clusters (e.g., \citealp{ferraro+09, ferraro+15},
\citealp{dalessandro+13_n362, simunovic+14}). The signature of these
processes, in fact, can remain imprinted in some BSS observational
features. This is the core concept of the so-called {\it dynamical
  clock} (see \citep[][hereafter F12]{ferraro+12} and subsequent
refinements in \citealt{alessandrini+16} and \citealt{lanzoni+16},
hereafter L16).

F12 demonstrated that the observed morphology of the BSS normalized
radial distribution\footnote{The ``normalized BSS distribution''
  \citep{ferraro+93} is defined as the ratio between the fraction of
  BSSs sampled in any adopted radial bin and the fraction of cluster
  light sampled in the same bin.}  (hereafter BSS-nRD) is shaped by
the action of dynamical friction (DF), which drives the objects more
massive than the average toward the cluster center, with an efficiency
that decreases with increasing radial distance. Thus, because of DF, a
peak followed by a minimum (a sort of BSS ``zone of avoidance''; see
also \citealp{mapelli04, mapelli06}) is expected to appear at small
radii in the BSS-nRD ratio and then propagate to larger and larger
distances from the cluster center as a function of time.  Accordingly,
F12 identified the observed shape of the BSS-nRD as a ``dynamical
clock'' able to measure the level of dynamical evolution experienced
by a cluster, and the radial location of the distribution minimum
($r_{\rm min}$) as the clock hand.  This feature, in fact, marks the
most external region of the system that has been significantly
affected by DF, and its distance from the center therefore depends on
how much a cluster is dynamically evolved.  This parameter, expressed
in units of the cluster core radius ($r_{\rm min}/r_c$), was used by
F12 to define families and sub-families of GCs in different stages of
their dynamical evolution (i.e., with different dynamical ages), all
having, however, the same same chronological age ($t\sim 12$-13 Gyr).

By using direct N-body simulations, \citet{alessandrini+16} proposed
an alternative and binning-independent parameter ($A^+$) to
characterize the progressive central segregation of BSSs. $A^+$ is
defined as the area enclosed between the cumulative radial
distribution of BSSs and that of a reference (and lighter) population.
L16 measured this quantity within one half-mass radius ($r_h$) for a
sample of GCs and found a well-defined correlation between $A^+$ and
the central relaxation time $t_{rc}$ of the system.  The new parameter
$A^+$ also strongly correlates with the hand of the dynamical clock,
$r_{\rm min}$ (see Figure 2 in L16). This is not obvious {\it a
  priori}, since the definitions of the two parameters are completely
independent. The evidence that, instead, they are mutually linked
through a tight and direct correlation indicates that they describe
the same phenomenon, i.e., the progressive central segregation of BSSs
due to the action of DF. These actually are different ways of
measuring the same process: as clusters get dynamically older, DF
progressively brings in BSSs from increasingly larger distances from
the center (thus generating a BSS-nRD minimum at increasingly larger
values of $r_{\rm min}$); correspondingly, BSSs accumulate toward the
cluster center (and the value of $A^+$ increases).  From an
operational point of view, $A^+$ provides some advantages with respect
to the method based on $r_{\rm min}$: $(i)$ $A^+$ is easier to
measure, since it does not require us to sample the entire radial
extension of the surveyed clusters; $(ii)$ it does not depend on
(somewhat arbitrary) assumptions about the radial binning; and $(iii)$
it describes a high-signal region where DF is accumulating BSSs
(instead of a low signal region, as the zone of avoidance).  The
latter property should also help us to reproduce the observations
through numerical simulations. Indeed, detecting the minimum of the
BSS-nRD is known to be a difficult task (e.g., \citealp{miocchi+15}),
since BSSs are an intrinsically scarce population and they rapidly
become less and less common in the external regions of a cluster. The
correct identification of $r_{\rm min}$ is then critically dependent
on the choice of the radial binning, since adopting intervals that are
too wide tends to wipe out the feature, while bins that are too narrow
generate noisy distributions because of the decreasing number of
sampled objects. Thus, it is not surprising that blind automatic
searches for $r_{\rm min}$ and its outward migration in numerical
simulations have been unsuccessful (e.g., \citealp{miocchi+15,
  hypki+17}). While this difficulty should not be confused with with a
lack of effectiveness for $r_{\rm min}$ to trace the dynamical
evolution of GCs, the parameter $A^+$ should simplify the
reproducibility of the observations via numerical simulations (which
already succeed in finding the central peak and the central BSS
segregation; \citealp{miocchi+15, hypki+17}).

As part of the HST UV Legacy Survey of GCs \citep{piotto+15} series,
this paper presents the determination of $A^+$ in 27 systems that have
been observed out to $\sim 1 r_h$ (see Sect. \ref{resu}). In
combination with the sample of L16, this provides us with a measure of
$A^+$ in roughly $32\%$ of the entire GC population in the Milky
Way. For this sample we find a strong correlation between $A^+$ and
the number of central relaxation times that have occurred in each cluster since
formation. A correlation between the level of BSS central segregation
(as quantified by $A^+$) and the degree of energy equipartition
experienced by BSSs in the cluster (as measured by the 
velocity-dispersion 
ratio of BSSs and MS-TO stars) is also found for the 14 GCs
in common with \citet{baldwin+16}.  The paper is organized as follows:
in Section \ref{obs} we present the UV approach to the study of BSSs
and the photometric database used; in Section \ref{bss} we describe
the BSS selection criteria; in Section \ref{resu} we determine $A^+$
and discuss the results.
 
\section{The photometric database and data analysis}
\label{obs}
\subsection{The UV approach}
\label{UVapproach}
Since the optical emission of old stellar systems is dominated by cool
bright giants, the systematic acquisition of complete samples of BSSs
at optical wavelengths is an intrinsically difficult task even with
HST \citep{ferraro+97, ferraro+99, ferraro+15}.  Instead, BSS searches
are particularly effective in the UV, where these stars appear among
the brightest objects in a GC and red giant branch (RGB) stars are
particularly faint. Thus, the usual problems associated with
photometric blends and crowding in the high-density central regions of
GCs are minimized, and BSSs can be reliably recognized and easily
separated from both other evolved and unevolved populations at UV
wavelengths.  Based on these considerations, several years ago we
first promoted the so-called {\it UV route to the study of BSSs in
  GCs} (see \citealp{ferraro+97, ferraro+99, ferraro+01, ferraro+03}),
an approach that allowed us to derive complete samples of BSSs even in
the central regions of the densest systems (see \citealp{lanzoni+07a,
  lanzoni+07b, lanzoni+07c, dalessandro+08, dalessandro+09, sanna+12,
  sanna+14, contreras+12, parada+16}).  Now, the dataset acquired in
the HST UV Legacy Survey of GCs (GO-13297; \citealp{piotto+15}) allows
us to extend this approach to a significant number of additional
clusters.  As a first step, in \citet[][hereafter, R17]{raso+17} we
presented the BSS populations obtained from these data in four GCs
(namely, NGC 2808, NGC 6388, NGC 6541, NGC 7078), and compared them to
the optical selections previously published for these same systems,
clearly demonstrating the great advantage of a {\it UV-guided} search
for BSSs relative to {\it optical-guided} hunts.  The discussion of
the BSS properties in terms of luminosity function, population ratios,
and correlations with the parent cluster properties for the entire
sample of GCs observed in the survey will be presented and discussed
in a forthcoming paper (Ferraro et al. 2018, in preparation).  Here we
focus on the 27 systems for which the observations cover radial
distances out to $\sim 1r_h$ (see Sect. \ref{resu}).

\subsection{The photometric database}
\label{database}
An overview of the HST UV Legacy Survey of GCs is presented in
\citet{piotto+15}\footnote{http://groups.dfa.unipd.it/ESPG/treasury.php}.
Here we summarize a few characteristics that are particularly relevant
for the study of BSSs.  Several images have been obtained for each
cluster in the F275W, F336W and F438W bands with the UVIS channel of
the WFC3.  WFC3/UVIS consists of two chips, each of $4096 \times 2051$
pixels, with a pixel scale of $0.04\arcsec$, resulting in a total
field of view of $\sim 162\arcsec \times 162\arcsec$.  In each band,
different pointings were dithered by several pixels, and in some cases
they were also rotated by $\sim 90^\circ$, to allow an optimal
subtraction of CCD defects, artifacts, charge loss and false
detections. All the images have been corrected for the effect of poor
charge transfer efficiency following \citet{anderson+10}.  In the
following we briefly describe the main novelty of the {\it UV-guided}
photometric approach here adopted (see also R17), with respect to the
early data releases based on an {\it optical-guided} reduction of the
images (see, e.g., \citealt{soto+17}).
 
The photometric catalogs were obtained using the software described by
\citet{anderson+08} and adapted to WFC3 images. Briefly, for each
image we obtained an ad-hoc array of PSFs by perturbing the tabulated
static PSFs\footnote{http://www.stsci.edu/$\sim$jayander/STDPSFs/}, to
properly take into account both the spatial and the temporal PSF
variations.  To extract the photometric catalogs from each individual
exposure by using the adopted arrays of PSFs we then ran the software
\texttt{img2xym$\_$wfc3uv}, which has been optimized for UVIS/WFC3
data by Jay Anderson and is similar to the \texttt{img2xym$\_$WFC}
program \citep{anderson+08}.  After correcting the stellar positions
for geometric distortion \citep{bellinibedin09, bellini+11}, we
determined the transformations between the single-exposure catalogs.
Finally, to take full advantage of the reduced crowding conditions at
UV wavelengths, we chose to perform the finding procedure on the F275W
and F336W images (see also R17) and then measured stars in all the
individual exposures. To this end, we used the \texttt{FORTRAN}
program \texttt{kitchen$\_$sync}, described by \citet{anderson+08} and
adapted to WFC3 images.  The final product is an astro-photometric
catalog containing the positions and the F275W, F336W, and F438W
magnitudes of all the stars found.  For a first-guess differential
reddening correction and field decontamination, information about
reddening and proper motions have been added to all the stars in
common with the catalogs published in the intermediate
release\footnote{Since these latter are based on an {\it
    optical-guided} data reduction, reddening and proper motion
  measures are available only for a sub-sample of the stars in our
  final catalogs.} \citep{piotto+15} and available upon request at the
  web page \texttt{http://groups.dfa.unipd.it/ESPG/treasury.php}.

\section{The BSS selection in UV-CMDs }
\label{bss}
R17 identified the purely UV ($m_{\rm F275W}, m_{\rm F275W}-m_{\rm
  F336W}$) CMD as the ideal diagram for BSS selection, since this
population is clearly distinguishable from the other evolutionary
sequences (see their Figure 1).  In particular, (1) along with the
hottest extension of the horizontal branch (HB), BSSs appear to be the
brightest objects in this CMD, while most of the RGB stars are
significantly fainter; (2) BSSs define a clean sequence populating a
strip that has a vertical extension of $\sim 3$ magnitudes and is
$\sim 2$ magnitudes wide in color; (3) BSSs are clearly
distinguishable from MS-TO and sub-giant branch (SGB) stars.

\begin{figure}[h!]
\centering \includegraphics[scale=0.8]{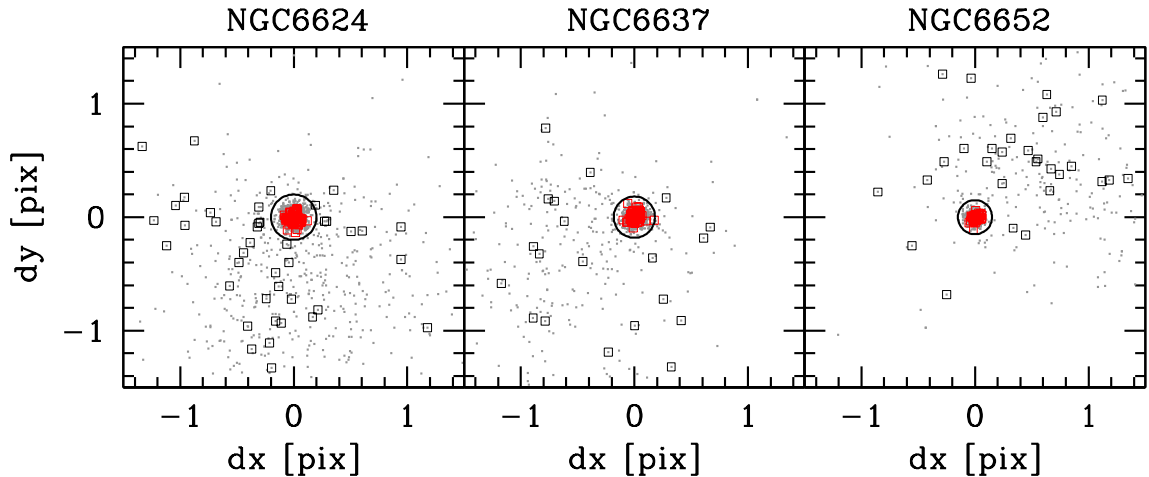}
\caption{Vector-point diagrams (VPDs) for three clusters in the survey
  (see labels), showing the measured displacements (in ACS/WFC pixels)
  of stars brighter than the MS-TO level (with $m_{\rm F275W}^*<0.5$),
  over a $\sim$7-8 yr time-baseline.  All stars within the circle (as
  well as stars with no displacement information) are assumed to be
  cluster members and have been included in the analysis, while those
  beyond the circle are considered as field contaminants. The red and
  black squares mark the cluster-member and the field-contaminant
  BSSs, respectively.}
\label{fig_pm}
\end{figure}

\begin{figure}[h!]
\centering \includegraphics[scale=0.8]{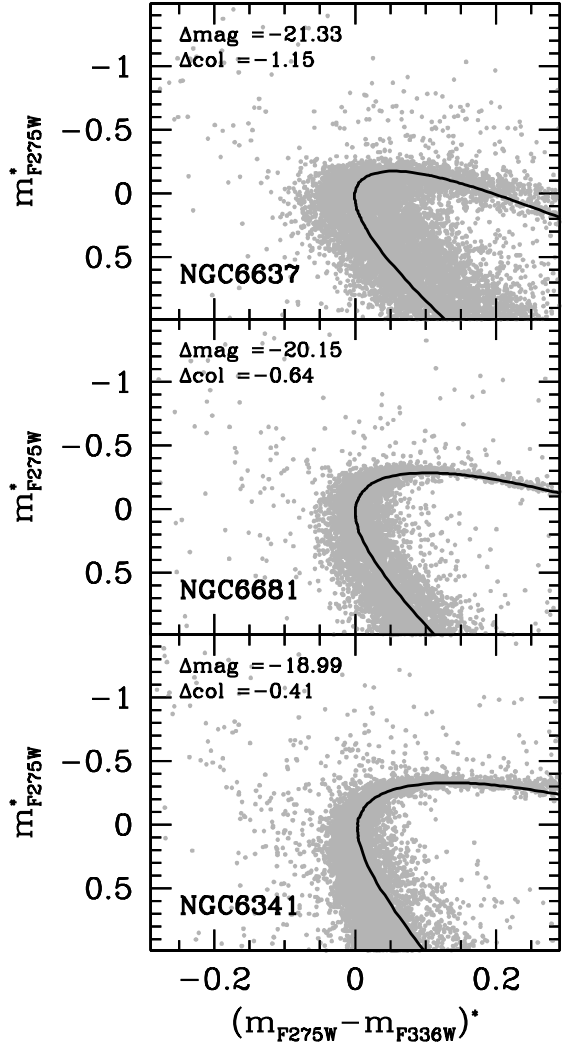}
\caption{Normalized CMD zoomed in the MS-TO region for three clusters
  (namely NGC 6637, NGC 6681, NGC 6341) belonging to the three
  metallicity groups defined in the text (from top to bottom, metal
  rich, metal intermediate and metal poor). The adopted 12-Gyr
  isocrone (from the BaSTI database) is shown as a solid line, and 
  shifts in color and magnitude have been adopted to locate the 
  MS-TO at $m_{\rm F275W}^*=0$ and $(m_{\rm F275W} -m_{\rm F336W})^*=0$ 
  are labelled in each panel. }
  \label{n_CMD}
\end{figure}

To perform a homogeneous selection of BSSs in clusters with different
values of distance, reddening and metallicity, we adopted the
procedure suggested by R17, defining a ``normalized'' CMD (hereafter
n-CMD) where the magnitudes and the colors of all the measured stars
in a given cluster are arbitrarily shifted to locate the MS-TO at
$m_{\rm F275W}^*=0$ and $(m_{\rm F275W} -m_{\rm F336W})^*=0$.  Since
the morphology of the MS-TO and the SGB region changes as a function
of metallicity, and because the surveyed GCs have quite different iron
abundances (ranging from [Fe/H]$=-0.4$ for NGC 6624, to [Fe/H]$=-2.4$
for NGC 7078), we adopted the following procedure to determine the
necessary shifts for a proper normalization of the CMDs:
\begin{enumerate}
\item While the updated determination of precise proper motions for
  all the clusters in the survey is in progress, here we used the
  currently available measurements for a first-order identification of
  field stars in the most contaminated systems.  Proper motions have
  been determined over a $\sim$7-8 yr time-baseline using the Advanced
  Camera for Surveys (ACS) Globular Cluster Survey data (GO-10775;
  \citealp{sarajedini+07}) as first-epoch observations and the new
  WFC3 UV positions as the second epoch.  To separate cluster members from
  field stars, we built the vector point diagrams (VPDs) plotting all
  the available displacement measures in each cluster. We thus used an
  iterative $\sigma$-rejection procedure\footnote{Starting from a
    first-guess circle of radius $r_i$ in the VPD, we selected all the
    enclosed stars and determined their dispersion along the two axes
    ($dx$ and $dy$). At the following step, the new selection radius
    is defined as $r=4\times \sigma_{xy}$, where
    $\sigma_{xy}=\sqrt{dx^2+dy^2}$. The final selection radius is then
    obtained when convergence is reached (typically, after five
    iterations).} to distinguish cluster members, which are
  well-grouped around the VPD center, from field contaminants,
  which generally exhibit a much more scattered distribution (see also
  \citealp{king+98, bedin+03, bellini+09, bellini+14, milone+12,
    massari+15, cadelano+17, soto+17}).  For illustration purposes we
  show in Figure \ref{fig_pm} the VPDs of three clusters. All stars
  with no displacement information have been conservatively retained as members in
  the following analysis.
\item All the stars for which we have color-excess determinations available were
  then corrected for differential reddening following the approach
  described, e.g., in \citet{milone+12, milone+17}.
\item The available clusters were divided into three groups according
  to their metallicity: metal-poor, with [Fe/H]$<-2$,
  metal-intermediate, with $-2<$[Fe/H]$<-1$, and metal-rich, with
  [Fe/H]$>-1$.
\item For each group, a 12 Gyr-old isochrone of appropriate
  metallicity was adopted from the BASTI database
  \citep{pietrinferni+06} and was plotted for reference in the n-CMD,
  i.e., in the CMD with the MS-TO located at (0,0).
\item The stellar magnitudes and colors in each cluster were then
  shifted to match the appropriate reference isochrone.
\end{enumerate}
For the sake of illustration, Figure \ref{n_CMD} shows the n-CMD zoomed in
the MS-TO region for three clusters representative of the three
adopted metallicity groups (see labels). The observed sequences at the
MS-TO/SGB level nicely agree with the shape of the appropriate
isochrone.

\begin{figure}[h!]
\centering \includegraphics[scale=0.75]{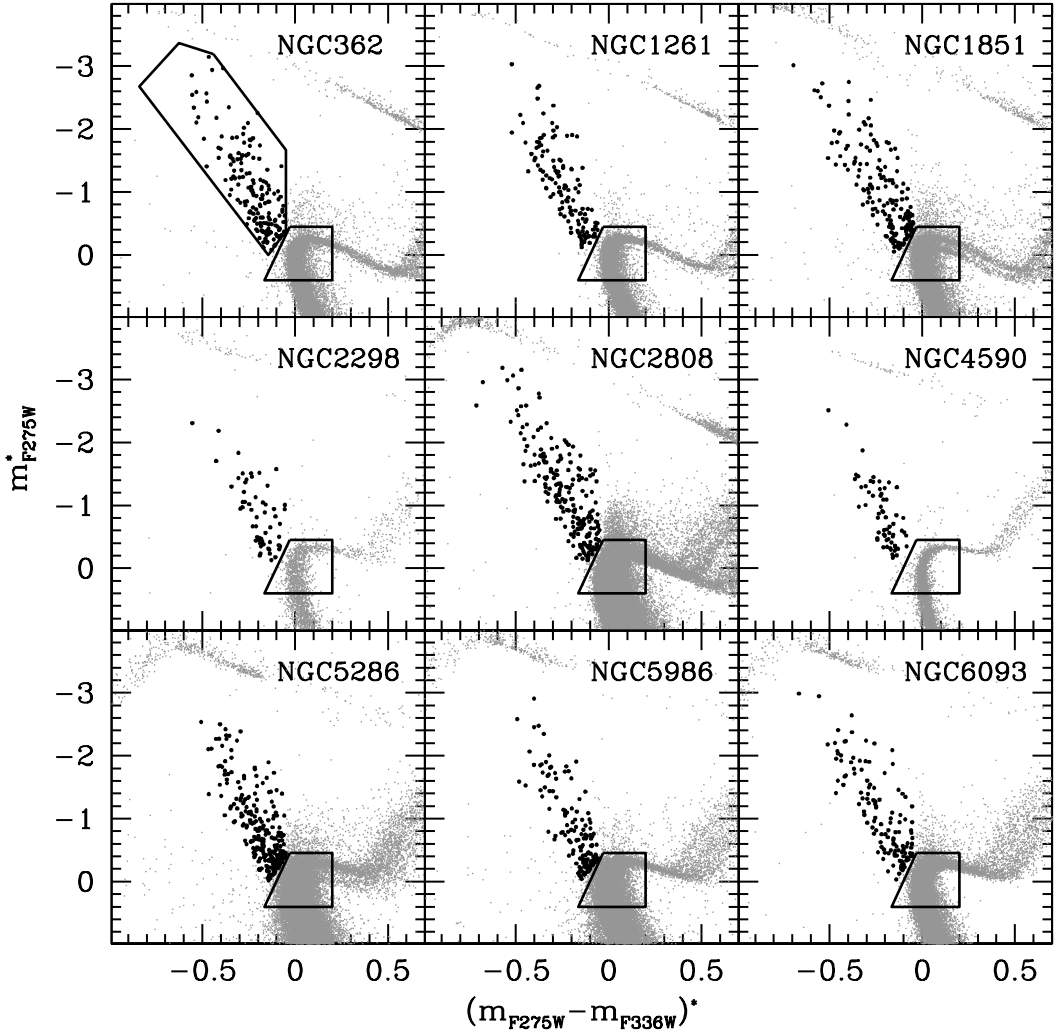}
\caption{The sample of BSSs (black dots) identified in each cluster is
  shown in the n-CMD.  The BSS selection box is drawn in the
    first panel, the one adopted for the MS-TO population is marked for
    all clusters.}
\label{cmd1}
\end{figure}

\begin{figure}[h!]
\centering \includegraphics[scale=0.75]{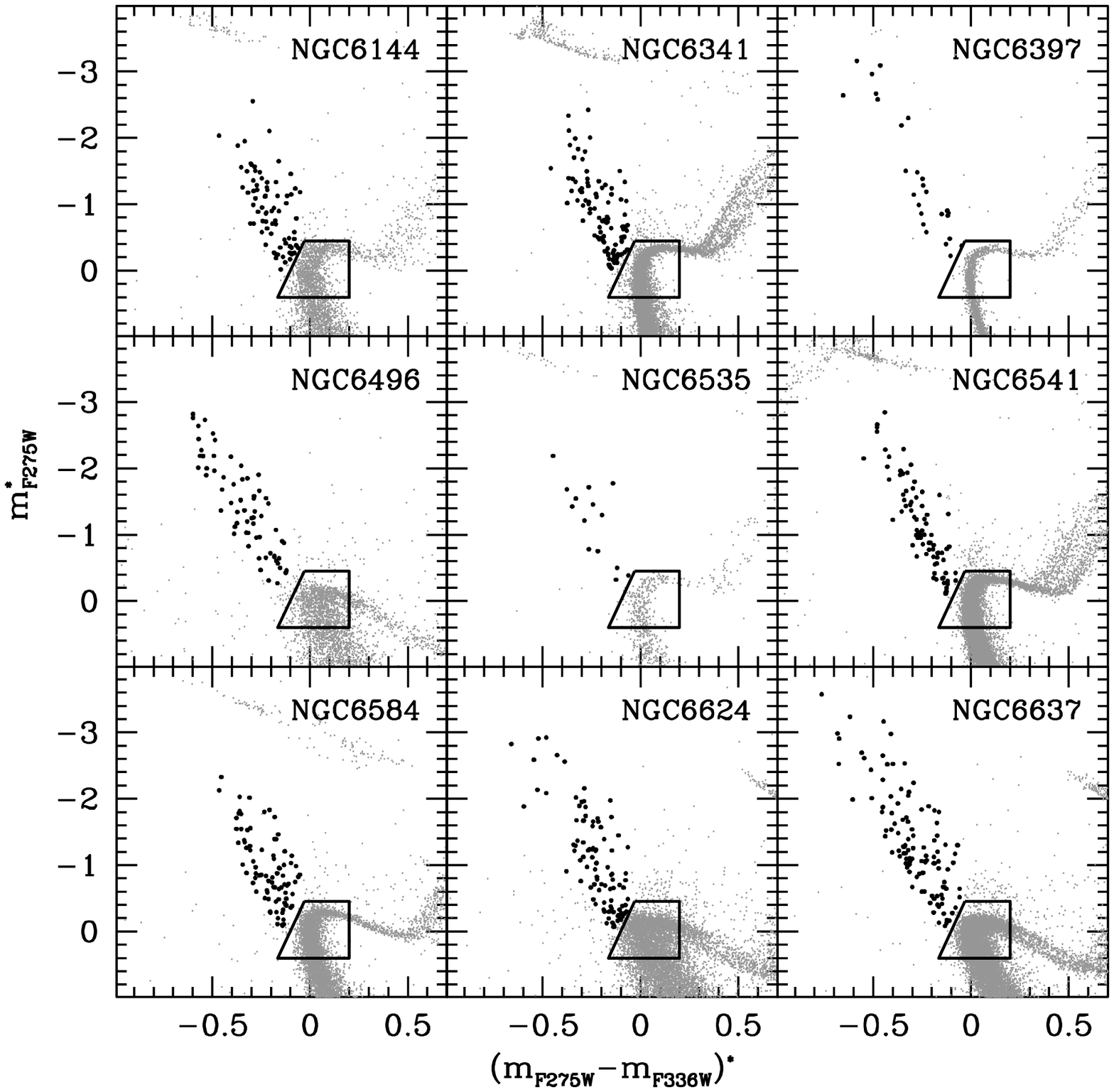}
\caption{As in Figure \ref{cmd1}.}
\label{cmd2}
\end{figure}

\begin{figure}[h!]
\centering \includegraphics[scale=0.8]{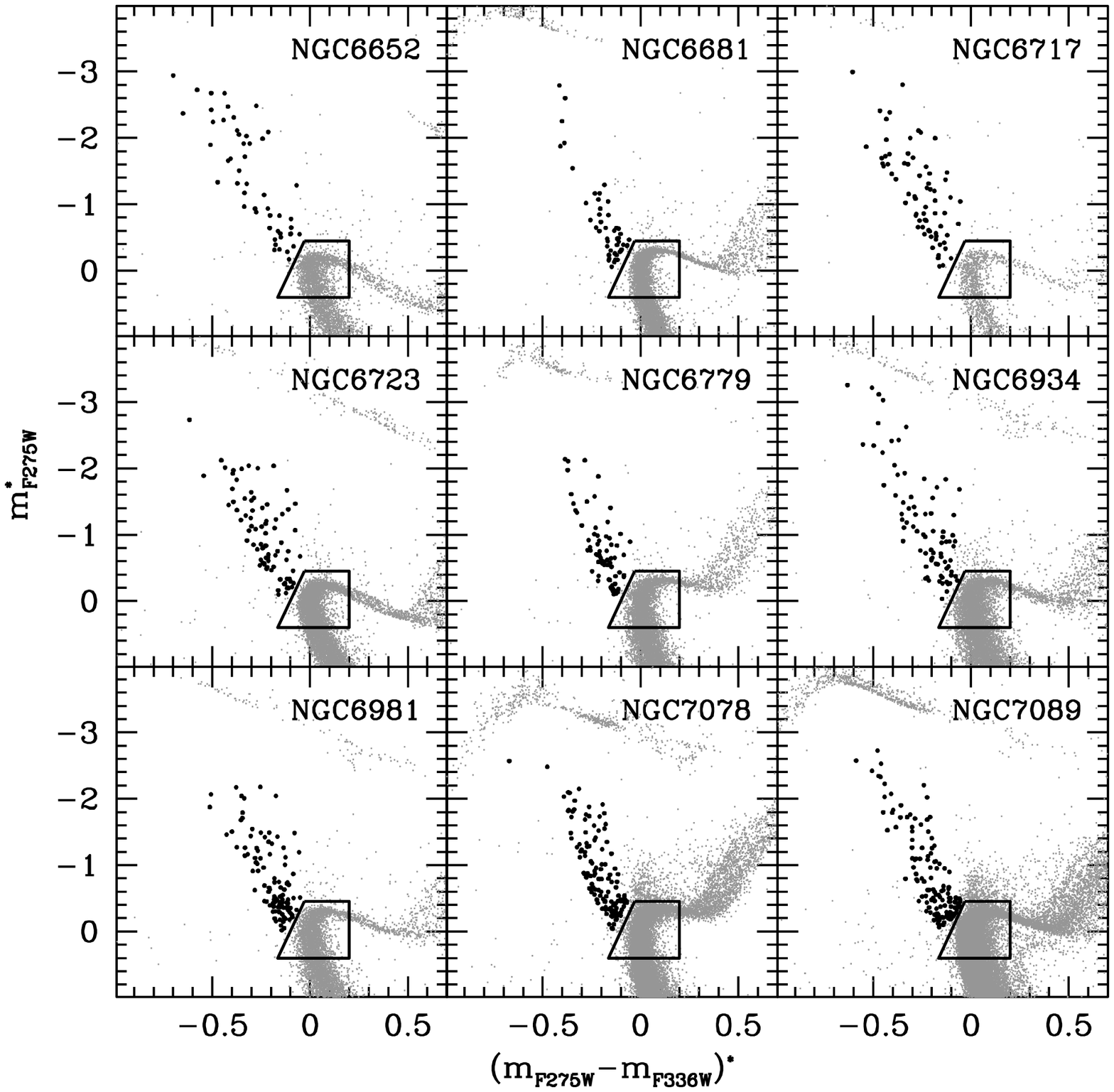}
\caption{As in Figure \ref{cmd1}.}
\label{cmd3}
\end{figure}

By construction, BSSs are expected to populate the same region in the
n-CMDs, Hence they can be selected in a homogeneous way by defining a
unique selection box  (see the upper-left panel in Figure 3).  Here we adopt the same boundaries defined in
R17, which have been designed to include the bulk of the BSS
population in all the clusters, independent of their metallicity (note
that the BSS sequence slightly moves from blue to red for increasing
iron content).\footnote{ For sake of completeness here we report
    the equations of the lines delimiting the BSS box (see the first
    panel in Figure \ref{cmd1}). The two parallel strips, designed
    to include the bulk of the BSS population independently of the
    cluster metallicity, have equations: $m_{\rm F275W}^*= 3.86 \times
    (m_{\rm F275W} -m_{\rm F336W})^* -1.48$ and $m_{\rm F275W}^*= 3.86
    \times (m_{\rm F275W}-m_{\rm F336W})^* +0.56$. The red boundary,
    located at $(m_{\rm F275W} -m_{\rm F336W})* = -0.05$, is needed to
    exclude the objects populating the ``plume'' above the MS-TO,
    which is visible in the most massive high-density clusters and is
    essentially due to photometric blends.  The bottom edge, which
    separates faint BSSs from MS-TO stars, is set at more than
    5-$\sigma$ from the typical mean color of the MS-TO distribution
    and has equation: $m_{\rm F275W}^*= -4 \times (m_{\rm F275W}
    -m_{\rm F336W})^* -0.58$.  Note that in some cases a few objects
    lying close to the selection box boundaries have been excluded to
    minimize the presence of contaminating stars. This has no impact
    on the discussed results.}  As discussed in R17, the bright edge
of the BSS selection box is needed to distinguish very luminous BSSs
from stars populating the blue portion of the HB (when present). Since
the HB morphology is expected to vary (primarily) as a function of the
metallicity, different bright boundaries of the BSS selection boxes
were adopted for the three metallicity groups.  Figures
\ref{cmd1}-\ref{cmd3} show the BSS sample selected in the 27 GCs
discussed in the paper.

For meaningful conclusions, the radial distribution of BSSs needs to
be compared with that of a population of normal cluster stars tracing
the overall density profile of the system (hereafter, REF population).
In order to measure the $A^+$ parameter as accurately as possible, we
choose to use MS stars around the MS-TO level.  The adopted REF
population selection box (see Figures \ref{cmd1}-\ref{cmd3}) is
delimited in magnitude between $m^*_{\rm F275W}=-0.45$ and $m^*_{\rm
  F275W}= 0.4$ and in color between $(m_{\rm F275W} - m_{\rm
  F336W})^*=0.2$ and the line $m^*_{\rm F275W} = -6.30\times (m_{\rm
  F275W} - m_{\rm F336W})^*-0.62$.  This portion of the CMD turns out
to provide the ideal REF population, since it includes several hundred
stars and therefore is negligibly affected by statistical
fluctuations.  It is also expected to be poorly affected by the
  possible presence of binary systems, since the number of single
  MS-TO stars is largely dominant within the adopted box.

\section{Results and Discussion}
\label{resu}
To measure the level of BSS segregation in the surveyed clusters we
used the parameter $A^+$, which is defined (see
\citealp{alessandrini+16}) as the area enclosed between the cumulative
radial distribution of BSSs, $\phi_{\rm BSS}(x)$, and that of a
reference (lighter) population, $\phi_{\rm REF}(x)$:
\begin{equation} A^+(x) = \int_{x_0}^x
  \phi_{\rm BSS}(x') -\phi_{\rm REF}(x') ~dx', \label{eq_A+}
\label{eq_apiu} 
\end{equation}
where $x_0$ and $x$ are the innermost and the outermost distances from
the cluster center considered in the analysis.  As discussed in
previous papers, the parameter $A^+$ has been designed to describe the
sedimentation-type process that progressively makes BSSs sink toward the
cluster center, thus tracing the level of dynamical evolution of the
system.  In principle, then, the measure of $A^+$ requires only that
the observations cover the cluster region most sensitive to the BSS
segregation (i.e. the central region), and that this be large enough
to allow a sufficient sampling of the cumulative radial distributions.
Moreover, to compare values of $A^+$ obtained in GCs with different
structures and sizes, this parameter should be determined within
equivalent radial portions in each system.  Determining the exact
extension of this portion is a complex task, since it depends on the
properties of the cluster and of its stellar populations (for
instance, the number of BSSs found within one core radius from the
center can be a very small or a relatively large number, depending on
the cluster).  Following L16 we measure $A^+$ out to one half-mass
radius (hereafter, $A^+_{rh}$) and we express distances on a logarithmic
scale to enhance the sensitivity of the parameter to the segregation
of BSSs, which is predominant in the central regions of each system
(hence $x = \log (r/r_h)$ in the above equation).

\begin{figure}[h!]
\centering \includegraphics[scale=0.75]{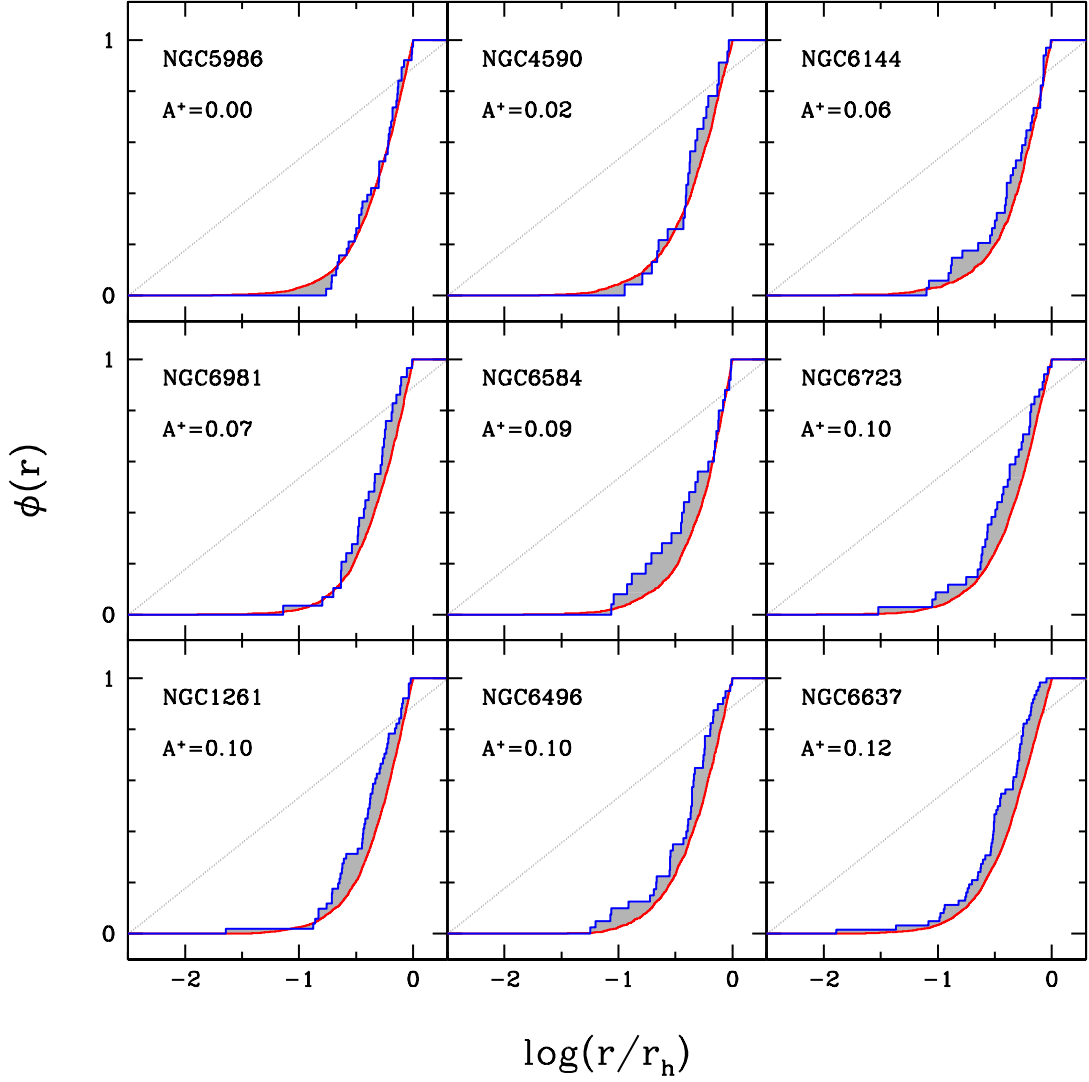}
\caption{Cumulative radial distributions of BSSs (blue line) and REF
  stars (red line) in the nine GCs with the smallest values of
  $A^+_{rh}$.  By construction (see Sect. \ref{bss}), the cumulative
  radial distributions are normalized to unity at $r_h$.  The size of
  the area between the two curves (shaded in grey) corresponds to the
  labelled value of $A^+_{rh}$ (see also Table 1). Clusters are ranked
  in terms of increasing value of $A^+_{rh}$.}
\label{fig_cumu1}
\end{figure}

\begin{figure}[h!]
\centering \includegraphics[scale=0.8]{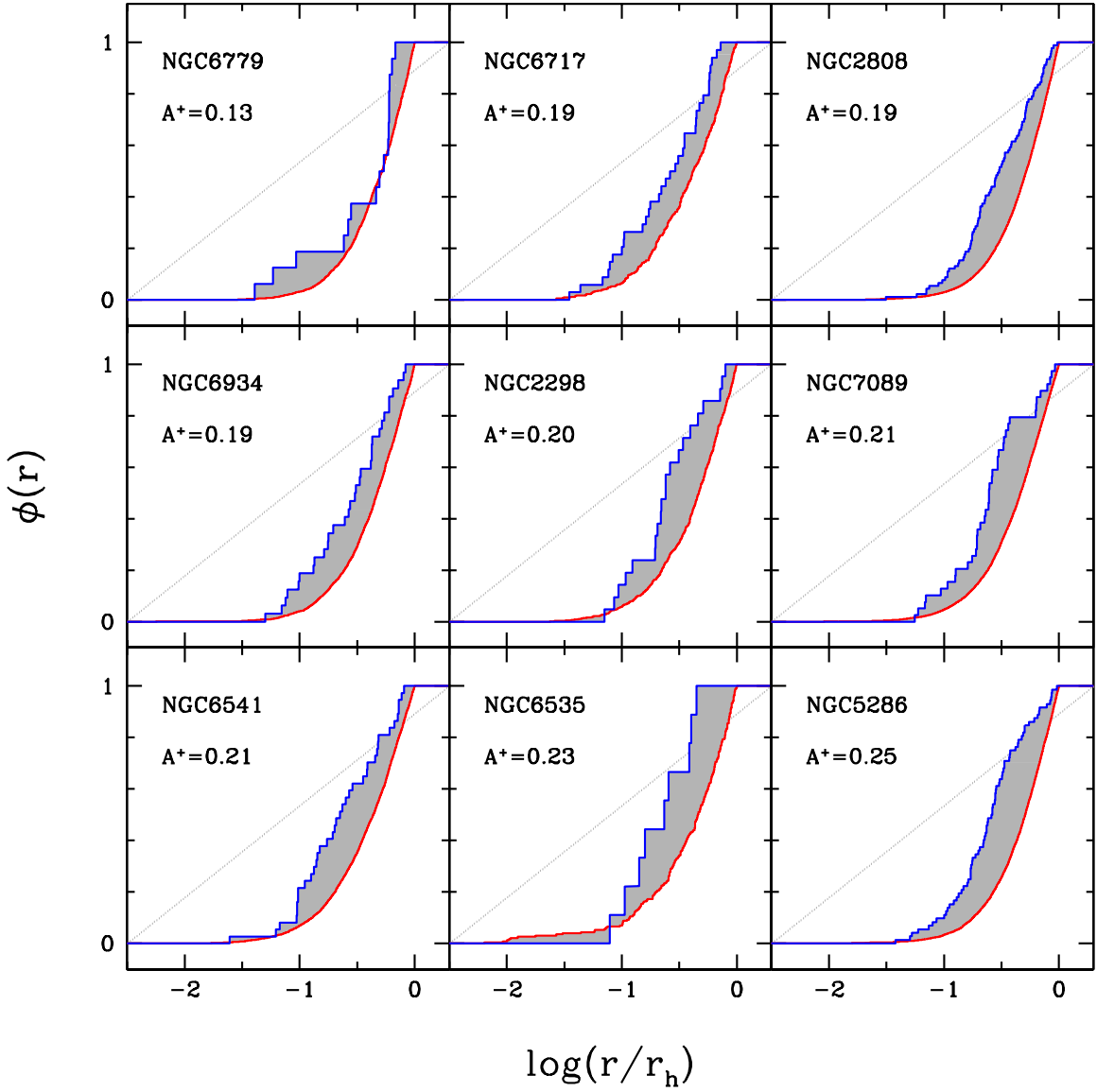}
\caption{As in Figure \ref{fig_cumu1}, for the nine GCs with
  increasingly larger value of $A^+_{rh}$.  }
\label{fig_cumu2}
\end{figure}

\begin{figure}[h!]
\centering \includegraphics[scale=0.8]{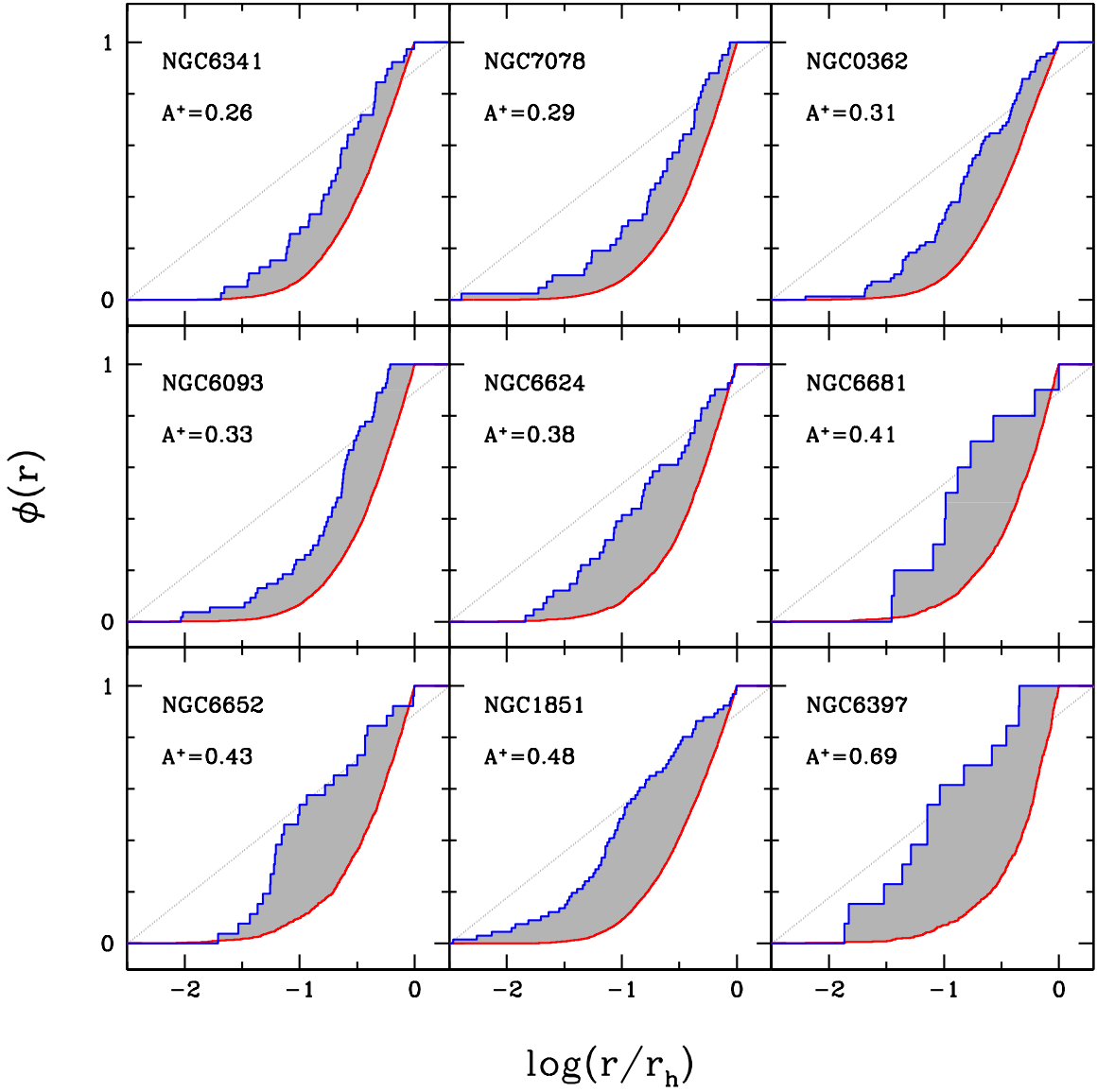}
\caption{As in Figure \ref{fig_cumu1}, for the remaining nine GCs,
  those with the largest values of $A^+_{rh}$.  }
\label{fig_cumu3}
\end{figure}

Our HST-WFC3 data covers approximately the innermost
$80\arcsec$-$85\arcsec$ in each surveyed cluster. This is large enough
to sample the entire half-mass radius in 22 of the clusters.  We also
added to the sample four clusters (namely, NGC 4590, NGC 6144, NGC
6496 and NGC 6723) for which the available data cover between $\sim
90$\% and 98\% of $r_h$.  We also added NGC 6397, which is known to be
a post-core collapse cluster and therefore is expected to belong to
the family of dynamically-old GCs (Family III as defined in F12) and
which has a BSS population highly segregated in the central regions.
Our data sample $\sim 30$ core radii in this cluster, and based on the
n-RDs discussed in F12, this is a large enough distance for a proper
study of BSS segregation. In fact, Figures 2 and 3 in F12 clearly show
that the region where the BSS-nRD shows a central peak (which is the
signature of BSS sedimentation) is of the order of a few core radii in
all clusters. Hence, the total sample extracted from the HST UV Legacy
Survey of GCs for the present analysis amounts to 27 objects.  To
increase the size of the sample, we also take into account 21 GCs from
L16.  The parameter $A^+$ in L16 has been determined from a
combination of UV HST observations and complementary wide-field
optical data from the ground able to sample the entire half-mass
radius region of each system.  To minimize the risk of stellar blends
mimicking BSSs at optical wavelengths (see an example in Figures 5 and
6 in \citealp{lanzoni+07b}), in L16 only the brightest portion of the
BSS distribution has been taken into account.  Hence, for consistency,
in the present analysis we consider only BSSs with $m^*_{\rm
  F275W}<-1.0$.  This selection corresponds to including only the most
massive tail of the BSS population, thus maximizing the sensitivity of
the $A^+$ parameter to the mass-segregation effect.  Although caution
is needed in deriving BSS masses from their luminosity distribution
(see \citealt{geller+12}), the adopted magnitude cut roughly
corresponds to selecting BSSs that are more massive than $\sim 1.2
M_\odot$.  The cumulative radial distributions of BSS and REF
populations here determined for the 27 investigated clusters are shown
in Figures \ref{fig_cumu1}-\ref{fig_cumu3}, and the corresponding
values of $A^+_{rh}$ are labeled in each panel and listed in Table
\ref{tab_apiu}.  The primary source of uncertainty on $A^+$ is the
relatively small-number statistics of the BSS sample in each system.
To estimate the errors on $A^+$ (see Table \ref{tab_apiu}) we
therefore used a jackknife bootstrapping technique
\citep{lupton93}.\footnote{For sake of consistency, we re-evaluated
  the errors on $A^+$ of the L16 sample using the same technique.}

Since the adopted BSS and REF populations have different magnitude
limits, they could be characterized by different levels of
completeness, which, in turn, could impact the value of $A^+_{rh}$.
Hence, we performed artificial star experiments to estimate the
photometric completeness of the two populations in some of the most
massive (hence most crowded) clusters with different central
densities.  In the worst cases of the intermediate density systems (as
NGC 2808) the completeness of the BSS and REF populations is always
larger than $\sim 90\%$, with the only exception being the faintest MS-TO
stars at $r<15\arcsec$, for which it decreases to $\sim 80\%$.  In the
most compact clusters (as M15) the selected populations are complete
at more than 90\% for $r>30\arcsec$, while the completeness decreases
to 80\% and 70-75\% at the faintest boundary of the MS-TO selection
box for $15\arcsec<r<30\arcsec$ and $r<15\arcsec$, respectively.  We
redetermined the value of $A^+_{rh}$ from 1000 random realizations of
the two population samples corrected for the estimated incompleteness
levels. The mean value and dispersion of the resulting $A^+_{rh}$
distributions are well within the uncertainties estimated with the
jackknife bootstrapping technique, thus indicating that the
photometric incompleteness of the adopted samples does not affect the
present analysis.

To investigate the connection between the BSS segregation level and
the dynamical status of the parent cluster, we studied the relation
between the measured values of $A^+_{rh}$ and the dynamical age of the
system quantified by the number of current central relaxation times
($t_{rc}$) that have occurred since the epoch of cluster formation
($t_{\rm GC}$): $N_{\rm relax} = t_{\rm GC}/t_{rc}$.  Because all
Galactic GCs have approximately the same age, this is simply an
alternative way of illustrating the connection between $A^+$ and the
dynamical evolution experienced by a cluster, with respect to the
direct comparison between $A^+$ and $t_{rc}$ adopted in previous works
(F12, L16 and R17).  Indeed, to keep the sources of uncertainty at
a minimum, for all the program clusters we assumed the same average age
($t_{\rm GC} = 12$ Gyr from the compilation of \citealt{forbes+10}).
Note however that adopting individual age estimates for each cluster
does not change the result. The central relaxation times have been
empirically estimated as in equation (10) of \citet{djorg93}, adopting
$0.3 M_\odot$ as average stellar mass, $M/L_V=2$ as $V$-band
mass-to-light ratio, the integrated $V$ magnitudes listed in
\citet{harris96}, and the structural parameters listed in Table
\ref{tab_apiu}.

\begin{figure}[h!]
\centering \includegraphics[height=15.0cm, width=15.0cm]{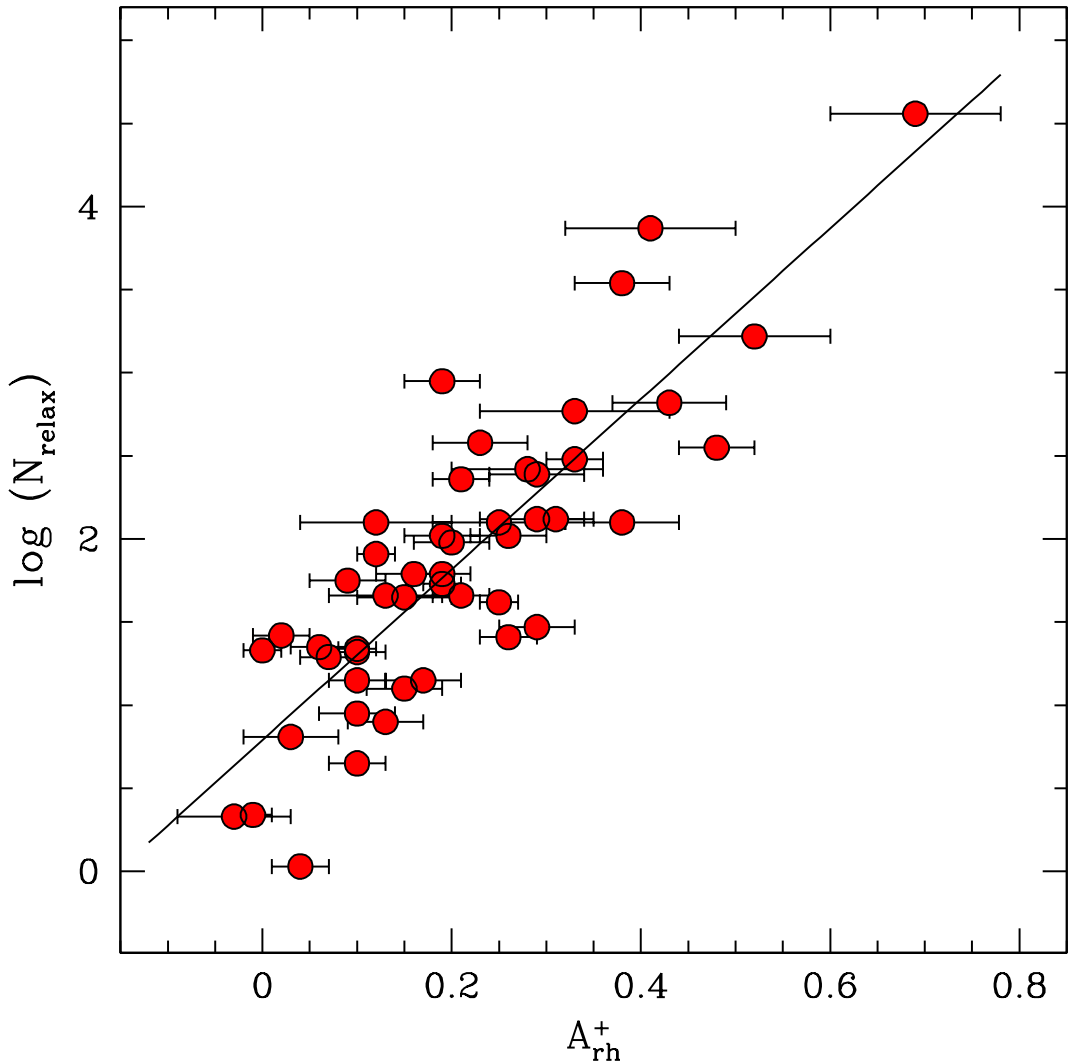}
\caption{Relation between $A^+_{rh} $ and $\log(N_{\rm relax})$ for
  the entire sample of 48 GCs.  The parameter $N_{\rm relax} = t_{\rm
    GC}/t_{rc}$ quantifies the number of current central relaxation
  times occurred since cluster formation. The tight relation between
  these two parameters demonstrates that the segregation level of BSSs
  measured by $A^+$ can be used to evaluate the level of dynamical
  evolution experienced by the parent cluster. }
\label{apiu_age}
\end{figure}

The nice correlation in Figure \ref{apiu_age} clearly shows that $A^+$
is a powerful indicator of cluster dynamical evolution.  The most
dynamically evolved cluster in the sample is NGC 6397, and the five
objects with the next largest values of $A^+_{rh}$ (namely, M30, NGC
1851, NGC 6652, NGC 6681, and NGC 6624) are all post-core collapse or
high-density systems.  The present study also increased the number of
clusters showing a modest level of dynamical evolution. In fact we
count at least 6-7 GCs with extremely low values of $A^+_{rh}$
($<0.05$-0.06).  Interestingly, we also find similarly small values
for other objects not included in the present study (as NGC 5053 and
NGC 6809) because the WFC3 field of view does not entirely sample
their half-mass radius.  This suggests that they may also belong to
this class, although more radially extended observations are
needed. Hence, a significant fraction (possibly $\sim 10\%$) of the
Galactic GC population could still be relatively unevolved
dynamically.

The best-fit relation to the observed points plotted in Figure
\ref{apiu_age} is:
\begin{equation}
  \log N_{\rm relax} = 5.1 (\pm 0.5) \times A^+_{rh} +0.79 (\pm 0.12)
\end{equation}
with a scatter of 0.47 and a high statistical significance: the
Spearman rank correlation coefficient is $\rho=0.82$,
and the Pearson correlation coefficient is $r=0.85$, indicating a
strong linear correlation between the two parameters. The relation
remains the same even if NGC 6397 which has the largest value of
$A^+_{rh}$, and/or the four GCs that are not sampled all the way out to
$r_h$ are excluded from the analysis.  On one hand, this new relation
is obtained from $\sim 1/3$ of the total GC population of the Milky
Way and thus definitely consolidates the idea that the segregation
level of BSSs can be used to evaluate the dynamical evolution
experienced by the parent cluster. On the other hand, the scatter of
this relation may indicate that further refinements should be used to
measure $A^+_{rh}$, and, more likely, that the empirical values of
$t_{rc}$ are crude approximations of the true relaxation times of
Galactic GCs (as discussed, e.g., by \citealp{chatterjee13_CC} from
dedicated Monte Carlo simulations). It is also worth noticing that the
current estimates of $t_{rc}$ are still largely based on cluster
parameters derived from surface brightness profiles, which could be
biased by the presence of few bright stars. In a future paper we will
present star count profiles for all the clusters in the HST UV Legacy
Survey of GCs, from which accurate values of the central density, core
and half-mass radii, and concentration parameter will be derived, and
new central relaxation times will be estimated. This might possibly
reduce the scatter and provide a refinement of the proposed relation.
The scatter notwithstanding, our analysis fully confirms that this is
the correct route for a proper description of the dynamical evolution
of star clusters.  The calibration of the relation via N-body or Monte
Carlo simulations requires that all the (known) ingredients (such as dark
remnants, primordial binaries, etc.) are taken into account. Indeed
preliminary N-body simulations \citep{alessandrini+16} have shown how
the inclusion of dark remnants can significantly change the BSS
segregation timescale in simulated clusters, hence caution should be
exercised when calibrating these observables via simulations of
simplified models.
 
\begin{figure}[h!]
\centering \includegraphics[height=15.0cm, width=15.0cm]{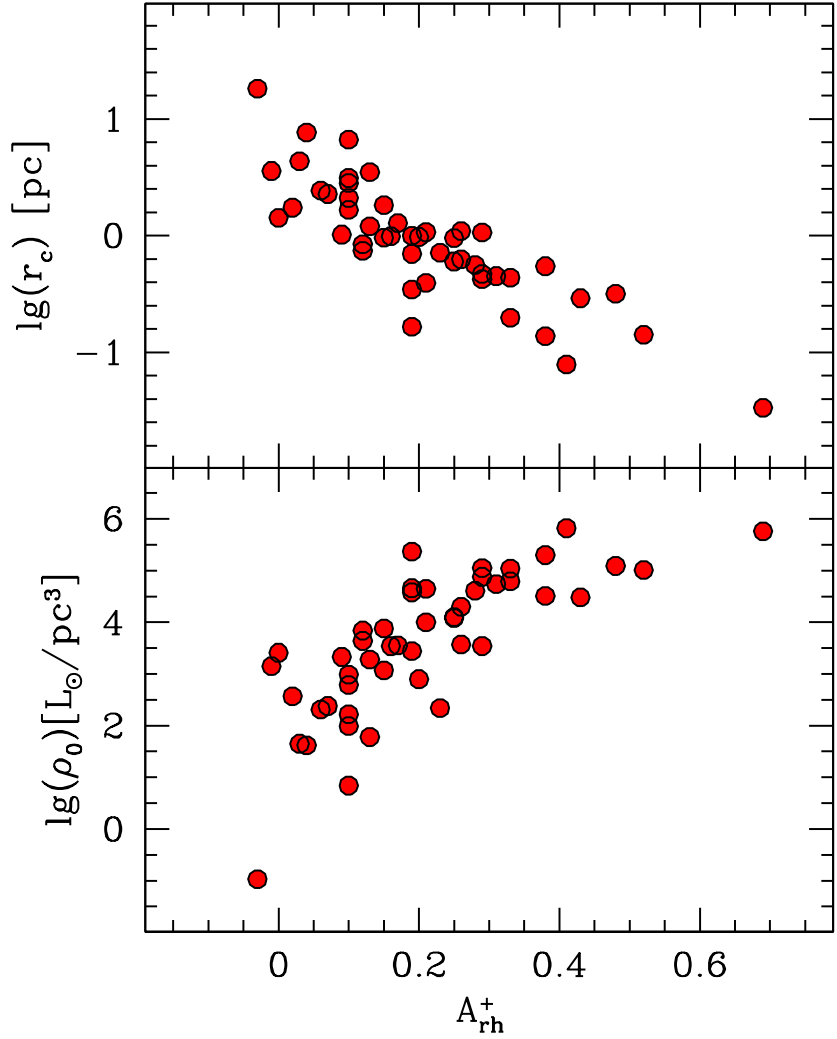}
\caption{ Relation between $A^+_{rh}$ and two physical parameters
    that are expected to change with the long-term dynamical evolution
    of GCs: the core radius {\it (upper panel)} and the central
    luminosity density {\it (lower panel)}, both taken from
    \citet{harris96}.}
\label{fig_para}
\end{figure}

 To further investigate the solidity of the $A^+$ parameter as
  dynamical aging indicator we also studied the dependence on
  $A^+$ of two physical parameters  that are expected to change with the long-term
  dynamical evolution of the cluster.  Figure \ref{fig_para} shows the behaviour of
  the core radius ($r_c$) and the central luminosity density ($\rho_0$, both  from
  \citealt{harris96}) as a function of $A^+$.  The well-defined trends shown
  in the figure, with $r_c$ decreasing and $\rho_0$ increasing with $A^+_{rh}$ (i.e., with increasing 
   dynamical age), nicely match  the expectations. The tight and strong detected relations, although 
   somehow predictable on the basis of the result shown in Figure \ref{apiu_age}, fully confirm
  the eligibility of the $A^+$ parameter as dynamical aging
  indicator.
 
\begin{figure}[h!]
\centering \includegraphics[height=15.0cm, width=15.0cm]{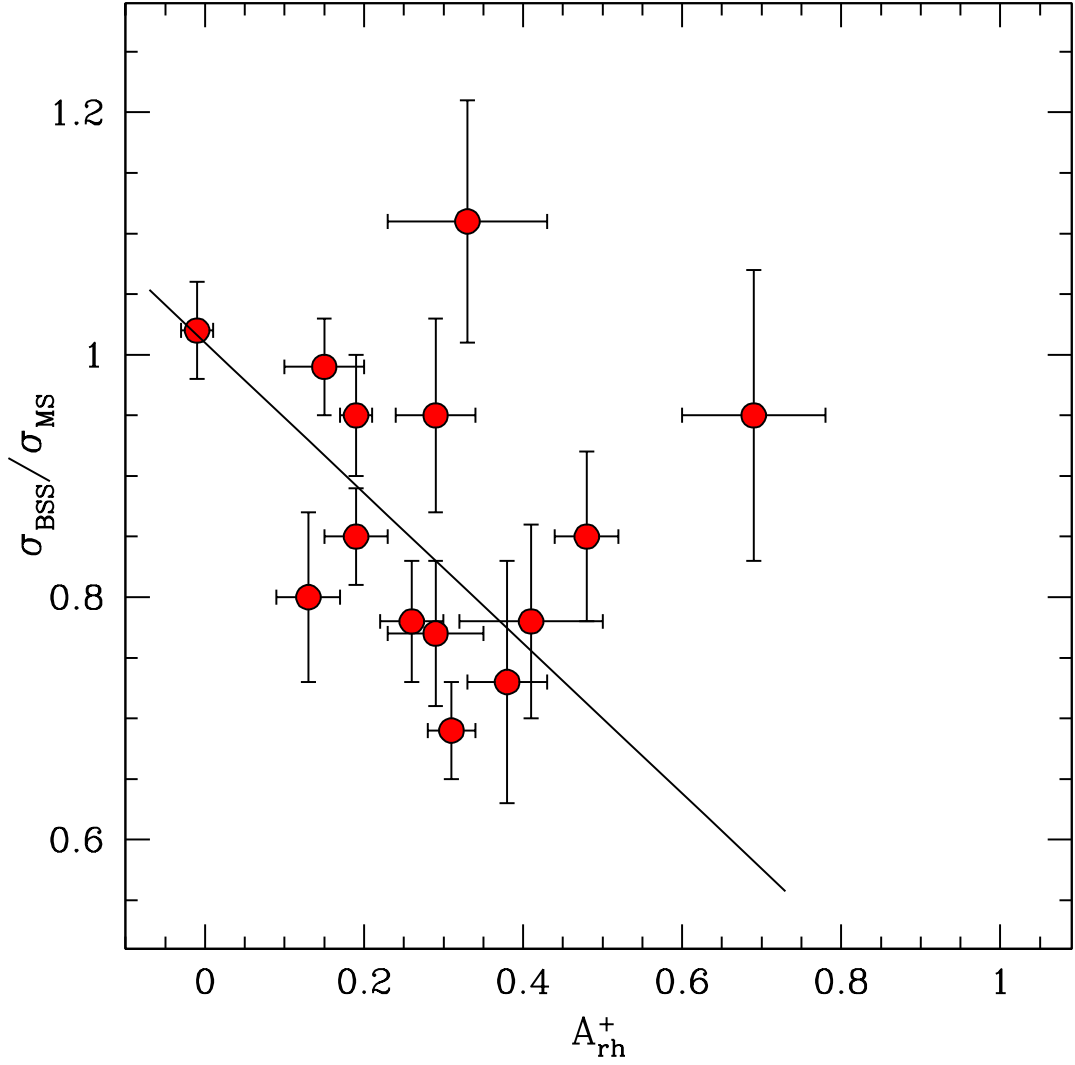}
\caption{Relation between $A^+_{rh}$ and the ratio $\alpha\equiv
  \sigma_{\rm BSS}/\sigma_{\rm MS-TO}$ between the BSS velocity
  dispersion and that of stars near the top of the MS for the 14 GCs
  in common with \citet{baldwin+16}.  The best-fit relation quoted in
  equation \ref{eq_alpha} is shown as a solid line.}
\label{fig_alpha}
\end{figure}

\citet{baldwin+16} measured the proper motions of 598 BSSs across 19
GCs, and inferred the ratio ($\alpha$) of the velocity dispersion of
BSSs ($\sigma_{\rm BSS}$) relative to that of stars near the top of
the MS ($\sigma_{\rm MS-TO}$), which is a measure of BSS equipartition
(see their equation 3).  Figure \ref{fig_alpha} compares this
parameter with our measurement of BSS mass segregation for the 14 GCs in
common between the two samples.  The best-fit relation is
\begin{equation}
     \alpha \equiv \frac{\sigma_{\rm BSS}}{\sigma_{\rm MS-TO}} = -0.62
     (\pm0.23) \times A^{+}_{rh} + 1.01 (\pm0.06)
\label{eq_alpha}     
\end{equation}
indicating a weak anti-correlation at $\sim 3\sigma$ confidence. The
size and the significance of the correlation are strengthened upon
omission of NGC 6397 (the cluster with the largest value of $A^+_{rh}$
in Figure \ref{fig_alpha}), which has the most uncertain velocity
dispersion in \citet{baldwin+16}, on account of their measuring 
only 10 BSSs. The Spearman
rank correlation coefficient is then $\rho = -0.39$, and the Person
correlation coefficient is $r = -0.38$. The relation indicates that
when $A^+_{rh} = 0$, then $\alpha \sim 1$.  So when there not yet been
sufficient time for mass segregation to develop, no energy
equipartition has developed yet either.  These results imply, as
expected, that as mass segregation develops, so does a certain level
of energy equipartition ($\sigma \propto M^{-\eta}$, where $M$ is the
stellar mass and $\eta$ trends over time from $0$ to a maximum value
of $\approx 0.1$-0.2, with $\eta = 0.5$ corresponding to complete
equipartition; \citealp{trenti+13}).  These results further confirm
our physical understanding of the dynamical evolution of GCs, as well
as our earlier arguments that $A^+$ can be used as the ``clock hand''
of a dynamical chronometer.

\acknowledgements{We thank the anonymous referee for useful comments
  that contributed to improve the presentation of the paper.
  FRF acknowledges the ESO Visitor Programme for the
  support and the warm hospitality at the ESO Headquarter in Garching
  (Germany) during the period when most of this work was
  performed. D.N, S.O. and G.P. acknowledge partial support by the
  Universit\`a degli Studi di Padova, Progetto di Ateneo CPDA141214
  ``Towards understanding complex star formation in Galactic globular
  clusters'' and by INAF under the program PRIN-INAF2014.  APM
  acknowledges support by the European Research Council through the
  ERC-StG 2016 project 716082 `GALFOR'.  JA and AB acknowledge the
  support of STScI Grant GO-13297. }

\newpage
\begin{table}[h!]
\begin{center}
\begin{tabular}{lcccccc}
\hline
 &  & & & \\
Name &  $c$ & $r_c$ & $r_h$ & $\log(t_{\rm rc})$ & $A^+_{rh}$ & $\epsilon_{A^+}$ \\
&  & & &  \\
\hline
NGC5986 & 1.23 &  28.2 &  58.8 & 8.75 &  -0.00 & 0.02 \\ 
NGC4590 & 1.41 &  34.8 &  90.6 & 8.66 &   0.02 & 0.03 \\ 
NGC6144 & 1.55 &  56.4 &  97.8 & 8.73 &   0.06 & 0.03 \\ 
NGC6981 & 1.21 &  27.6 &  55.8 & 8.79 &   0.07 & 0.03 \\ 
NGC6584 & 1.47 &  15.6 &  43.8 & 8.33 &   0.09 & 0.04 \\ 
NGC6723 & 1.11 &  49.8 &  91.8 & 8.93 &   0.10 & 0.03 \\ 
NGC1261 & 1.16 &  21.0 &  40.8 & 8.74 &   0.10 & 0.02 \\ 
NGC6496 & 1.18 &  35.6 &  93.6 & 8.76 &   0.10 & 0.03 \\ 
NGC6637 & 1.38 &  19.8 &  50.4 & 8.17 &   0.12 & 0.02 \\ 
NGC6779 & 1.38 &  26.4 &  66.0 & 8.42 &   0.13 & 0.06 \\ 
NGC6717 & 1.71 &   8.0 &  45.0 & 7.13 &   0.19 & 0.04 \\ 
NGC2808 & 1.56 &  15.0 &  48.0 & 8.35 &   0.19 & 0.02 \\ 
NGC6934 & 1.53 &  13.2 &  41.4 & 8.29 &   0.19 & 0.03 \\ 
NGC2298 & 1.38 &  18.6 &  58.8 & 8.10 &   0.20 & 0.04 \\ 
NGC7089 & 1.57 &  15.4 &  66.3 & 8.42 &   0.21 & 0.03 \\ 
NGC6541 & 1.86 &  10.8 &  63.6 & 7.72 &   0.21 & 0.03 \\ 
NGC6535 & 1.56 &  17.5 &  73.4 & 7.50 &   0.23 & 0.05 \\ 
NGC5286 & 1.41 &  16.8 &  43.8 & 8.46 &   0.25 & 0.02 \\ 
NGC6341 & 1.74 &  14.6 &  85.0 & 8.06 &   0.26 & 0.04 \\ 
NGC7078 & 2.29 &   8.4 &  60.0 & 7.69 &   0.29 & 0.05 \\ 
NGC0362 & 1.73 &  13.0 &  73.8 & 7.96 &   0.31 & 0.03 \\ 
NGC6093 & 1.74 &   7.0 &  40.6 & 7.60 &   0.33 & 0.03 \\ 
NGC6624 & 2.50 &   3.6 &  49.2 & 6.54 &   0.38 & 0.05 \\ 
NGC6681 & 2.50 &   1.8 &  42.6 & 6.21 &   0.41 & 0.09 \\ 
NGC6652 & 1.80 &   6.0 &  28.8 & 7.26 &   0.43 & 0.06 \\ 
NGC1851 & 1.95 &   5.4 &  51.0 & 7.53 &   0.48 & 0.04 \\ 
NGC6397 & 2.50 &   3.0 & 174.0 & 5.52 &   0.69 & 0.09 \\ 
\hline
\end{tabular} 
\end{center}
\caption{Structural/dynamical parameters and values of $A^+_{rh}$ for
  the 27 program clusters:  concentration parameter
  (column 2), core and half-mass radii in arcseconds (columns 3 and 4,
  respectively), logarithm of the central relaxation time in Gyr
  (column 5), derived value of $A^+_{rh}$ and its error (columns 6 and
  7). The structural parameters are from \citet{miocchi+13}, L16,
  \citet{cadelano+17}, and \citet{harris96} if not available in the
  previous studies, but for NGC 6717 and NGC 6535 and NGC 6496 for
  which we performed new determinations (Ferraro et al. 2018, in
  preparation). Clusters are ordered in terms of increasing value of
  $A^+_{rh}$.}
\label{tab_apiu}
\end{table}

\newpage

\newpage

\newpage

\newpage

\newpage

\newpage

\newpage

\newpage

\newpage

\newpage

\newpage

\end{document}